\documentclass[twocolumn,secnumarabic,amssymb,graphics,floatfix,nofootinbib,tightenlines,nobibnotes,aps,notitlepage,superscriptaddress,10pt]{revtex4-1}
\usepackage{graphicx}% Include figure files
\usepackage{dcolumn}% Align table columns on decimal point
\usepackage{bm}% bold math
\usepackage{bigints}
\usepackage{xcolor}  % for writting parts in color, e.g. in red

\newcommand{\ket}[1]{\ensuremath{| #1 \rangle}}

\newcommand{\be}{\begin{equation}}
\newcommand{\ee}{\end{equation}}
\newcommand{\ba}{\begin{array}}
\newcommand{\ea}{\end{array}}
\newcommand{\bqa}{\begin{eqnarray}}
\newcommand{\eqa}{\end{eqnarray}}
\newcommand{\bes}{\begin{equation}\begin{split}}
\newcommand{\ees}{\end{split}\end{equation}}
\usepackage{dcolumn}% Align table columns on decimal point
\usepackage{makecell} % for tables
\usepackage{blindtext}
\usepackage{newfloat}
\DeclareFloatingEnvironment[name={S}]{suppfigure}
\begin{document}
\title{Optimal control of non-Markovian dynamics in a single-mode cavity strongly coupled to an inhomogeneously broadened spin ensemble}

\author{Dmitry O. Krimer$^\S$}
\email[]{dmitry.krimer@gmail.com}
\affiliation{Institute for Theoretical Physics, Vienna University of Technology (TU Wien), 1040 Vienna, Austria, EU}
\author{Benedikt Hartl$^\S$}
\affiliation{Institute for Theoretical Physics, Vienna University of Technology (TU Wien), 1040 Vienna, Austria, EU}
\author{Florian Mintert}
\affiliation{Department of Physics, Imperial College, SW7 2AZ London, United Kingdom, EU}
\author{Stefan Rotter}
\affiliation{Institute for Theoretical Physics, Vienna University of Technology (TU Wien), 1040 Vienna, Austria, EU}
\let\thefootnote\relax\footnotetext{$^\S$These authors contributed equally to this work.}
\pacs{42.50.Ct, 42.50.Pq, 03.65.Yz, 42.50.Dv}

\begin{abstract}
Ensembles of quantum mechanical spins offer a promising platform for quantum memories, but proper functionality requires accurate control of unavoidable system imperfections. We present an efficient control scheme for a spin ensemble strongly coupled to a single-mode cavity based on a set of Volterra equations relying solely on weak classical control pulses. The viability of our approach is demonstrated in terms of explicit storage and readout sequences that will serve as a starting point towards the realization of more demanding full quantum mechanical optimal control schemes.
\end{abstract}

\maketitle

\section{Introduction}

In the past decade, we have witnessed tremendous progress in the implementation of elementary operations for quantum information processing. Single qubit gates can be realized with fidelities reaching $1-10^{-6}$ \cite{PhysRevLett.113.220501}, and also two-qubit gates can be implemented in a variety of systems \cite{nature.426.264,nature.447.836}.
With all these elements at hand, it is nowadays possible to implement quantum algorithms on architectures with a few qubits (on the order of five) \cite{PhysRevX.6.031007} and to engineer quantum metamaterials based on an ensemble of superconducting qubits coupled to a microwave cavity \cite{Macha:2014aa,{Shulga:2017}}. Implementing quantum logics on larger architectures, however, will most likely require a separation between quantum processing units and quantum memory units, where qubits in the former units admit fast gate operations and the qubits in the latter units offer long coherence times.

Since extended coherence times naturally imply weak interactions with other degrees of freedom, the sufficiently fast swapping of quantum information between processing and memory units is a challenging task. The most promising route to overcome slow swapping is the encoding of quantum information as a collective excitation in a large
ensemble composed of many ($N$) constituents, since this increases the swapping speed by a factor of $\sqrt{N}$. Among promising realizations of such ensembles those based on spins, atoms, ions or molecules are of particular interest \cite{Stanwix2010,Amsuess2011,Kubo2010,Probst2013,Rabl2006,Afzelius:2009,Verdu2009}. In many cases, however, system imperfections result in broadening effects giving rise to rapid dephasing of ensemble constituents - a restriction that limits the coherence times of such collective quantum memories.

As a result, various protocols to ensure the controlled and reversible temporal dynamics in the presence of inhomogeneous broadening were recently the subject of many studies. One of the proposed techniques in this context is the so-called controlled reversible inhomogeneous broadening (CRIB) approach \cite{Moiseev:2001aa,Kraus2006,Tittel:2010}, which is based on a rather subtle preparation method and on the inversion of atomic detunings during the temporal evolution. Most of the techniques developed for this purpose are based on photon-echo type approaches in cavity or cavity-less setups, such as those dealing with spin-refocusing \cite{PhysRevA.88.062324,PhysRevX.4.021049}, with atomic frequency combs (AFC) \cite{Gisin:2007,Staudt:2007aa,Riedmatten:2008,Jobez:2014,Riedmatten:2015,Moiseev2012} or with electromagnetically induced transparency (EIT) \cite{Novikova:2008}. Traditionally, these architectures operate in the optical region and require additional high-intensity control fields. The resulting large number of excitations is prone to spoil the delicate quantum information that is encoded in states with extremely low numbers of excitations. It would therefore be much better to work with low-intensity control fields, which, however, have the other problem to become easily correlated with the quantum memory. For the identification of control strategies, this implies that one may no longer treat the many different memory spins as independent objects, but that the (macroscopically) large ensemble needs to be described by a quantum many-body state. This makes any description of dynamics and an identification of control strategies a seemingly hopeless task.

In this paper we develop a very efficient semiclassical optimization technique based on a set of Volterra integral equations, which allows us to write information into a large, inhomogeneously broadened spin ensemble coupled to a single cavity mode by means of optimized classical microwave pulses and to retrieve it at some later time in the form of well separated cavity responses. In contrast to established echo techniques our scheme only involves low-intensity signals and therefore diminishes the influence of noise caused by writing and reading pulses. The applicability of our approach is also demonstrated in conjunction with a spectral hole-burning technique \cite{Krimer:2015aa,Putz:2017aa,Krimer:2016aa} that allows us to reach storage times going far beyond the dephasing time of the inhomogeneously broadened ensemble. Importantly, the Volterra equation exactly governs the resulting linear non-Markovian dynamics not only in the semiclassical but also in the pure quantum case for the particular situation without external drive, when all spins are initially in the ground state and the cavity contains initially a single photon \cite{Krimer:2015aa, Krimer:2016aa, Putz:2014aa}. Furthermore, the system's density function or nonequilibirum Green's functions, which show up in the framework of a full quantum-mechanical description, also satisfy mathematically similar integro-differential Volterra equations \cite{Lei2012aa, Zhang2012aa}. Hence, although the problem is treated semiclassically in what follows, we believe that our approach can be generalized to pure quantum regimes as well in which case the inclusion of the transient two-time correlation function of the cavity operator between the write and the readout may be needed -- an issue that will be postponed for future studies.

\begin{figure}[t!]
\includegraphics[width=0.475\textwidth]{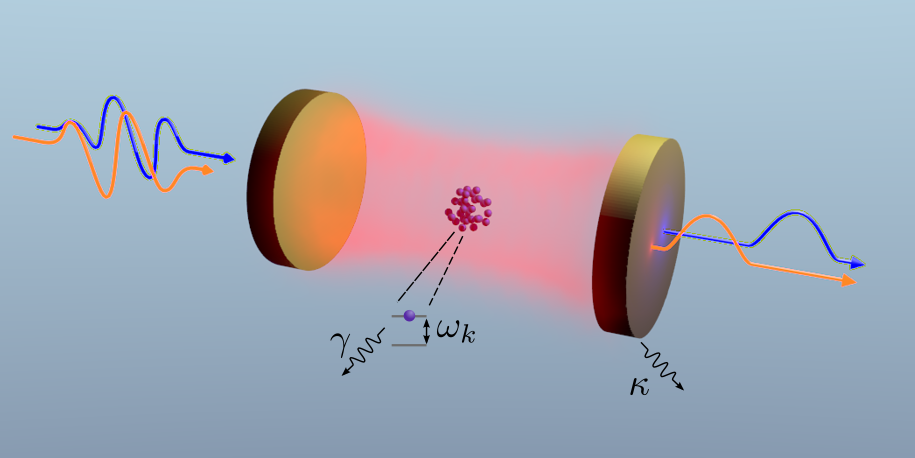}
\caption{Schematics of a single-mode cavity characterized by a frequency $\omega_c$ and a loss rate $\kappa$, coupled to an ensemble of two level atoms (spheres) with transition frequencies $\omega_k$ and a loss rate $\gamma \ll \kappa$. Curves designate optimized input and (non-overlapping) output signals.}
\label{fig:schematic}
\end{figure}
\begin{figure*}
\label{fig:sepzero:short:short:delay}
\includegraphics*[width=0.495\textwidth]{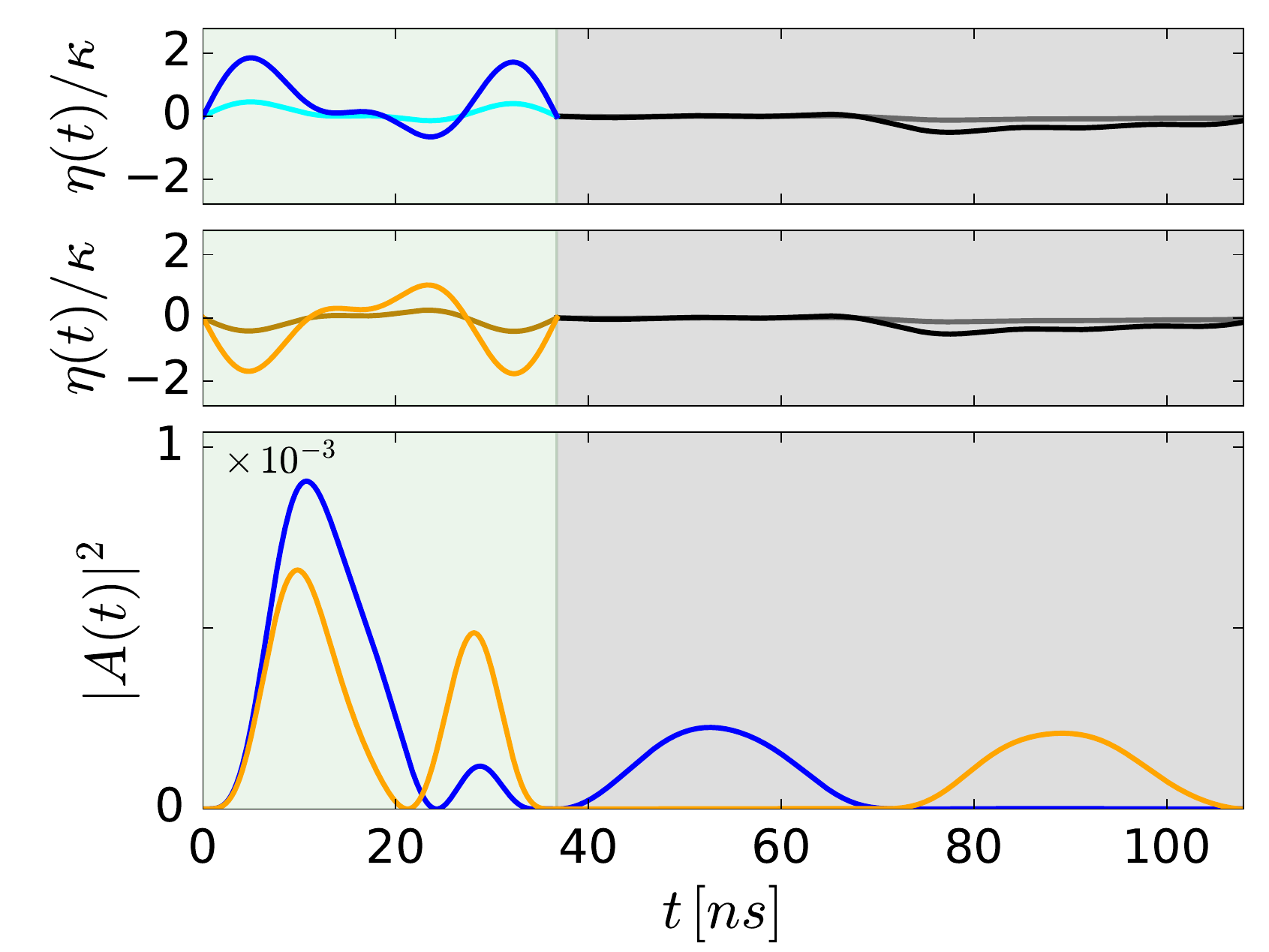}
\label{fig:sepzero:short:long:delay}
\includegraphics*[width=0.495\textwidth]{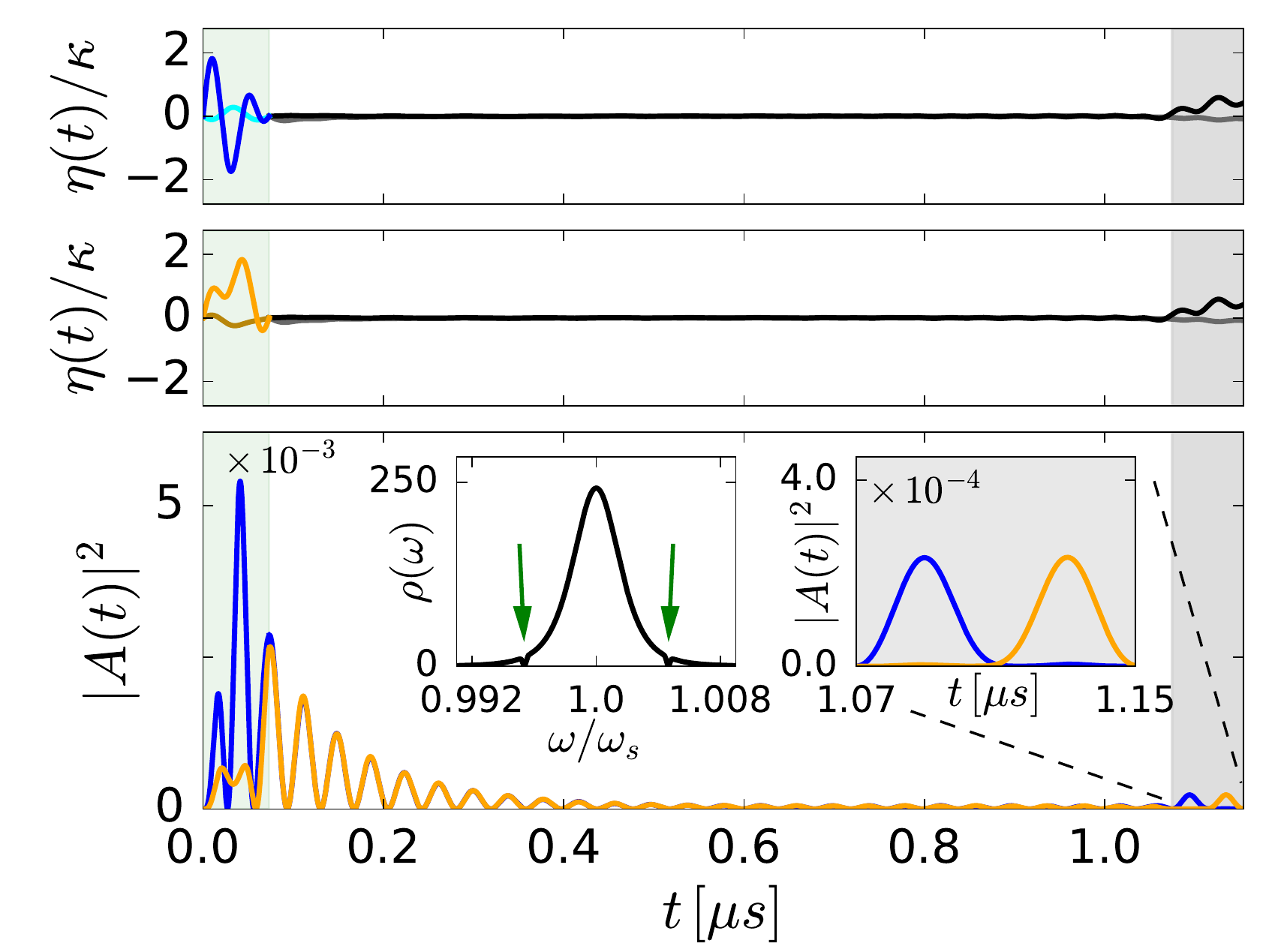}
\caption{Preparation of the spin ensemble configurations $\ket{0}$ and $\ket{1}$ for a spin density $\rho(\omega)=C\cdot\left[1-(1-q)(\omega-\omega_s)^2/\Delta^2\right]^{1/(1-q)}$ following a $q$-Gaussian distribution with $q=1.39$ centered around the cavity frequency $\omega_s=\omega_c$ and a full-width at half maximum $\gamma_q=2\Delta\sqrt{(2^q-2)/(2q-2)}=2\pi\cdot9.4$\,MHz. This form for $\rho(\omega)$ was established in our previous studies by a careful comparison with the experiment \cite{Putz:2014aa, KPMS14}. The right column shows that two holes were burnt into $\rho(\omega)$ at frequencies $\omega_s\pm\Omega$ (two arrows in the inset) to suppress decoherence \cite{Krimer:2015aa,Putz:2017aa} and to make room for a delay section [white area] between the write [green area (light gray)] and readout [gray area] sections. (In the inset $\rho(\omega)$ is plotted in units of $\omega_s^{-1}$.) The top and middle rows show real [blue (dark gray) and orange (light gray), respectively] and imaginary parts [cyan (light gray) and brown (dark gray), respectively] of the optimized write pulse $\eta^{(W)}_{\ket{0/1}}(t)$ for state ${\ket{0/1}}$ and of the generic readout pulse $\eta^{(R)}(t)$ (black and gray).The bottom row shows the cavity probability amplitude squared $|A(t)|^2$ for the resulting nonoverlapping cavity responses $A_{\ket{0}}^{(R)}(t)$ [blue (dark gray)] and $A_{\ket{1}}^{(R)}(t)$ [orange (light gray)]. The carrier frequency of all pulses, $\omega_p=\omega_c=2\pi\cdot 2.6915$\,GHz, and the coupling strength $\Omega/2\pi=12.5\,$MHz. The ratio of the powers between the readout and write pulses is 0.068 (0.013) for the case without (with) hole burning. The amplitudes of all pulses (two upper rows) are presented in units of $\kappa/2\pi=0.4\,$MHz.}
\label{fig:sepzero:delayed}
\vspace{-0.2cm}
\end{figure*}
\section{Theoretical model}

To be specific, we consider an ensemble of spins strongly coupled to a single-mode cavity via magnetic or electric dipole interaction as sketched in Fig.~\ref{fig:schematic}. All typical parameter values are chosen here in accordance with the recent experiment \cite{Putz:2014aa}, the dynamics of which can be excellently described by the Tavis-Cummings Hamiltonian \cite{Tavis:1968aa} (in units of $\hbar$)
\begin{eqnarray}
{\cal H}&=&\omega_ca^{\dagger}a+\frac{1}{2}\sum_j^N\nolimits\omega_j\sigma_j^z+\text{i}\sum_j^N\nolimits\left[g_j\sigma_j^-a^{\dagger}-g_j^*\sigma_j^+a\right]-
\nonumber\\
&-&\text{i}\left[\eta(t) a^{\dagger}\text{e}^{-\text{i}\omega_p t}-\eta(t)^* a\text{e}^{\text{i}\omega_p t}\right]\,.
\label{Hamilt_fun}
\end{eqnarray}
Here  $\sigma_j^{\pm},\,\sigma_j^z$ are the Pauli operators associated with each individual spin of frequency $\omega_j$ and $a^{\dag},\,a$ are creation and annihilation operators of the single cavity mode with frequency $\omega_c$.  An incoming signal is characterized by the carrier frequency $\omega_p$ and by the envelope $\eta(t)$. The interaction part of ${\cal H}$ is written in the dipole and rotating-wave approximation (terms $\propto a\sigma_j^-,\,a^\dag \sigma_j^+$ are neglected), where $g_j$ is  the coupling strength of the $j$-th spin. The distance between spins is assumed to be large enough such that the direct dipole-dipole interactions between spins can be neglected. Furthermore, the large number of spins allows us to enter the strong-coupling regime of cavity QED with the collective coupling strength, $\Omega=(\sum_j^Ng_j^2)^{1/2}$ \cite{Sandner:2012aa}, which leads to the enhancement  of a single coupling strength, $g_j$, by a factor of $\sqrt{N}$ ($N\approx 10^{12}$ in \cite{Putz:2014aa}).

We are aiming at the transfer of information from the cavity to the spin ensemble, its storage over a well-defined period of time, and its transfer back to the cavity. Our control scheme thus consists of a {\it write} and {\it readout section}, with a variable {\it delay section} in between. Starting from a polarized state with all spins in their ground state, we construct (i) two write pulses $\eta^{(W)}_{\ket{0}}(t)$ and $\eta^{(W)}_{\ket{1}}(t)$ that encode the respective logical states $\ket{0}$ and $\ket{1}$ in the spin ensemble. During the {\it delay section} (ii) the information is subject to dephasing by the inhomogeneous ensemble broadening and the external drive is optimized here to reduce the cavity amplitude $A(t)\equiv\langle a(t) \rangle$ (to prevent the information in the spin ensemble from leaking back to the cavity prematurely). In the {\it readout section} (iii) we switch on the readout pulse $\eta^{(R)}(t)$ (with substantially lower power than $\eta^{(W)}_{\ket{0/1}}(t)$) that maps the two logical states of the spin ensemble on two mutually orthogonal states of the cavity field, expressed by the cavity amplitude  $A^{(R)}_{\ket{0}}(t)$ or $A^{(R)}_{\ket{1}}(t)$ respectively. Note that the write pulses (i) are specific for the input states $\ket{0}$ and $\ket{1}$, but pulses (ii) and (iii) are generic as they are designed without prior knowledge of the information stored in the ensemble. (For the sake of simplicity we formally absorb the delay pulse into $\eta^{(R)}(t)$.) The goal of our work is to find optimal time-dependent choices for $\eta^{(W)}_{\ket{0}}(t)$, $\eta^{(W)}_{\ket{1}}(t)$ and $\eta^{(R)}(t)$, such that $A^{(R)}_{\ket{0}}(t)$ and $A^{(R)}_{\ket{1}}(t)$ have minimal temporal overlap in analogy to time-binned qubits where information is stored in the occupation amplitudes of two well distinguishable time bins \cite{Brendel:1999, Staudt:2007aa}.

We describe the dynamics by deriving the equations for the spin and cavity expectation values, $\langle \sigma_k^{-}(t)\rangle$ and $A(t)$, under the Holstein-Primakoff-approximation \cite{Primakoff1939} ($\langle \sigma_k^{(z)} \rangle \approx -1$) valid in the regime of weak driving powers (the number of the excited spins is always small compared to the ensemble size). This allows us to formally express $\langle \sigma_k^{-}(t)\rangle$ as a time integral with respect to $A(t)$ and to develop an efficient framework in terms of Volterra equations that relate cavity amplitudes $A(t)$ and pump profiles $\eta(t)$ \cite{KPMS14},
\be
A(t)=\int_0^t d\tau\ {\cal K}(t-\tau) A(\tau)+{\cal D}(t)\ ,
\ee
where ${\cal D}(t)$ depends on the time integral of the driving signal and on the initial conditions for the cavity amplitude as well as of the spin ensemble. The memory kernel function ${\cal K}(t-\tau)$, which is responsible for
the non-Markovian feedback of the spin ensemble on the cavity, is proportional to the collective coupling strength, $\Omega^2$, and explicitly depends on a spectral spin distribution characterized by a function $\rho(\omega)$ (see Appendix \ref{App_A}.) When switching on a constant drive, the system exhibits damped oscillations characterized by the Rabi frequency, $\Omega_R \approx 2\Omega$, and the total decoherence rate, $\Gamma$, mostly determined by the dephasing caused by the inhomogeneous broadening of the spin ensemble \cite{KPMS14}.

A consequence of the linearity of the governing Volterra equations is that for two pump profiles $\eta_{1/2}(t)$, resulting in the two cavity amplitudes $A_{1/2}(t)$, any coherent superposition of these pulses $c_1\eta_1(t)+c_2\eta_2(t)$ will result in the corresponding cavity amplitudes $c_1A_1(t)+c_2A_2(t)$.

The Volterra equation for the cavity amplitude is physically the classical correspondence of the Heisenberg cavity spin equations on the level of expectation averages after elimination of the spin ensemble variables (see Appendix \ref{App_A}). However, as was demonstrated in \cite{Krimer:2015aa, Putz:2014aa, Diniz:2011aa}, the Volterra equation also governs quantum spin--cavity dynamics for the particular case when all spins are initially in the ground state and the cavity contains initially a single photon. Therefore, we take the amplitude of the write pulses, $\eta^{(W)}_{\ket{0/1}}(t)$, such that the net power injected into the cavity corresponds to the power of a coherent driving signal with an amplitude equal to the cavity decay rate, $\kappa$. The latter prepares on average a single photon in the empty cavity for stationary transmission experiments (see Appendix \ref{App_D} for details). Due to the linearity of the Volterra equations, also rescaling their solutions by a global prefactor leaves them perfectly valid.
\section{Optimal control scheme}

As a first step we need to find optimal write and readout pulses which prepare the logical spin ensemble configurations $|0\rangle$ and $|1\rangle$ and map them onto well distinguishable cavity responses. We do this through the optimization of a functional, which ensures the minimal overlap between the cavity amplitudes $A_{\ket{0}}^{(R)}(t)$ and $A_{\ket{1}}^{(R)}(t)$ of the logical states $|0\rangle$ and $|1\rangle$ in the readout section by exploring various temporal shapes of both the write pulses, $\eta_{|0/1\rangle}^{(W)}(t)$, and of the readout pulse, $\eta^{(R)}(t)$. In practice we expand all involved driving pulses in a basis of trial functions $\sin(n\, \omega_f t)$ ($n=1,2,...$) with the fundamental frequency $\omega_f$ defined as the inverse of the time duration of the write or readout section counted in multiples of  half the Rabi period, $\pi/\Omega_R$. Next, we construct the functional defined as the time-overlap integral between $A_{\ket{0}}^{(R)}(t)$ and $A_{\ket{1}}^{(R)}(t)$ in the readout section. We then search the functional's minima under several constraints considering the expansion coefficients as unknown variables using the standard method of Lagrange multipliers (see Appendices \ref{App_A}, \ref{App_B}). Due to the linearity of governing equations with respect to the control pulses this procedure, as seen in Appendix \ref{App_B}, is numerically highly efficient since the time integration of the Volterra equations can be performed independently of the subsequent optimization of the expansion coefficients of the control pulses.

A typical result of this optimization (first without a delay section) is depicted in Fig.~\ref{fig:sepzero:delayed} (left column), where the amplitudes of all optimized pulses as well as those of the resulting cavity responses are depicted. One can indeed see that the two different configurations stored in the spin ensemble, $\ket{0}$ and $\ket{1}$, are retrieved by the same readout pulse in the form of two well-separated cavity responses. The storage efficiency can be quantified in terms of the ratio of integrated cavity amplitudes during the readout and write section, which turns out to be $\approx40\,\%$ for the configurations $\ket{0}$ and $\ket{1}$ shown in Fig.~\ref{fig:sepzero:delayed} (left column).

The bottleneck for extended information storage times in the ensemble is its inhomogeneous broadening, as determined by the continuous spectral density $\rho(\omega)$ appearing in our theoretical description. Specifically, the total decoherence rate in the limit of strong coupling (when $\Omega>\Gamma$) can be estimated as $\Gamma\approx\kappa+\pi\Omega^2\rho(\omega_s\pm\Omega)$ \cite{Diniz:2011aa,KPMS14}, indicating that the dominant contribution to $\Gamma$ stems from the spectral density $\rho(\omega)$ at frequencies close to the maxima of the two polaritonic peaks, $\omega=\omega_s\pm\Omega$. To suppress this decoherence rate $\Gamma$ it is thus advisable to work with spin ensembles having a spectral density that falls off faster than $1/\Omega^2$ in its tails such that $\Gamma\to \kappa$ for large $\Omega$. The corresponding ``cavity protection effect'' \cite{Diniz:2011aa,KPMS14,Kurucz:2011aa} has meanwhile been demonstrated also experimentally \cite{Putz:2014aa}, but has the drawback of requiring prohibitively large coupling strengths to take full effect. Alternatively, one can burn two narrow spectral holes at frequencies close to $\omega_s\pm\Omega$, during a preparatory step for $t\le0$. This technique \cite{Krimer:2015aa,Putz:2017aa,Krimer:2016aa} was recently shown to be both easily implementable and very efficient in suppressing the decoherence rate $\Gamma$ even below the bare cavity decay rate $\kappa$ \cite{Putz:2017aa}. Incorporating this hole burning protocol in the present analysis allows us to increase the dephasing time from $1/\Gamma \sim 75\,$ns [the case shown in Fig.~\ref{fig:sepzero:delayed} (left column)] to microsecond time scales [see Fig.~\ref{fig:sepzero:delayed} (right column)] for which we can now meaningfully introduce a delay section in between the write and the readout section. In Fig.~\ref{fig:sepzero:delayed} (right column) we show that with parameters taken from recent experiments \cite{Putz:2017aa} we can extend the storage time and thereby our method's temporal range of control beyond one micro-second. Evidently, such an extension of the storage time comes with a reduced efficiency which is here as large as $5\,\%$.
\begin{figure}
\hspace*{-0.32cm}
 \includegraphics[width=0.5\textwidth]{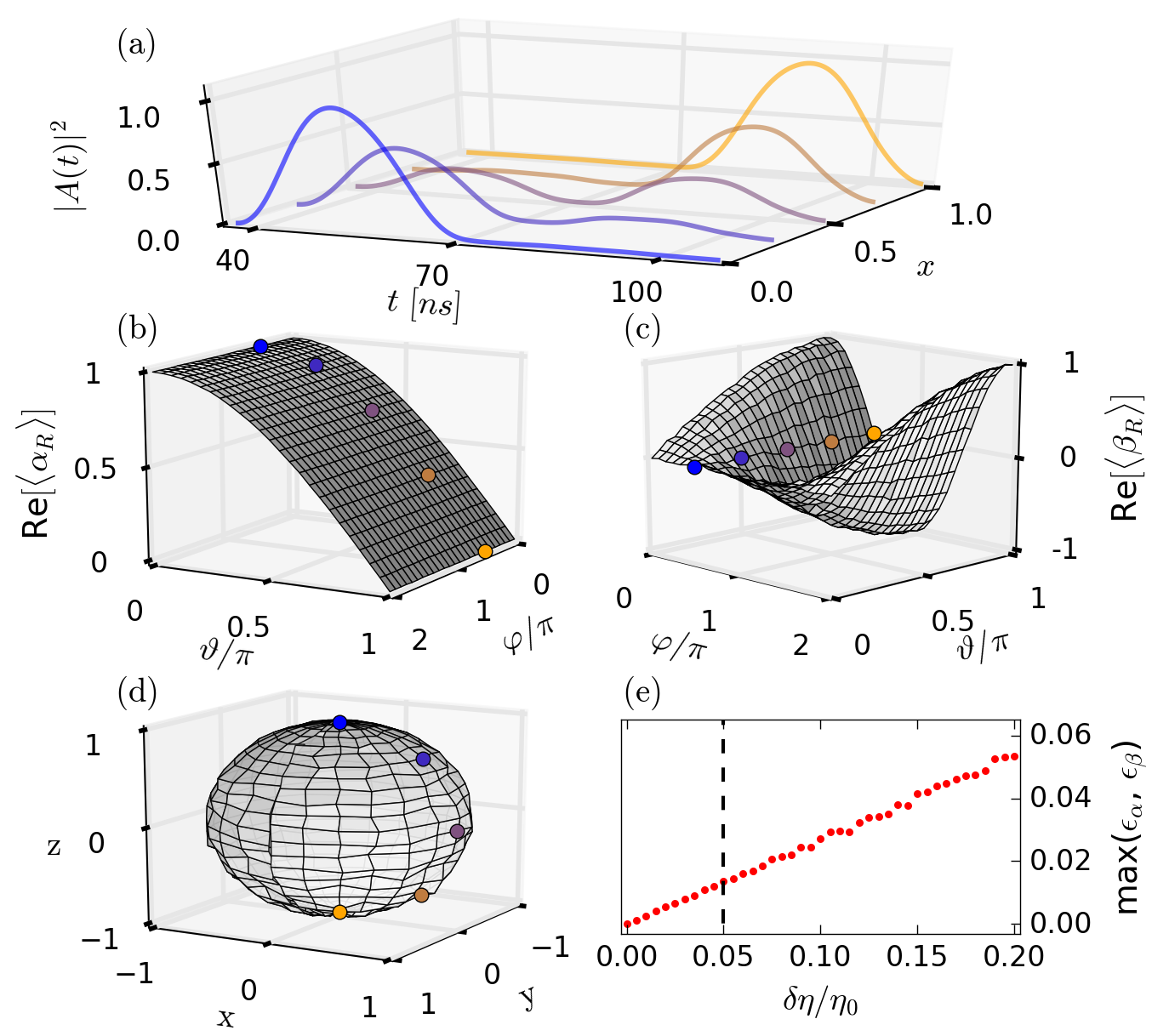}
\caption{{(a)} Retrieved cavity amplitude in the readout section, resulting from a superposition of write pulses, $\alpha_x \!\cdot \! A^{(R)}_{\ket{0}}(t)+\beta_x\! \cdot \! A^{(R)}_{\ket{1}}(t)$ (normalized to a maximum value of $1$) 	in the absence of noise with the write amplitude $\eta_0=\kappa$. We used the rebit parametrization $\alpha_x$ and $\beta_x$ (see text) from $x=0$ to $1$ in steps of $0.25$ (parameters as in Fig.~\ref{fig:sepzero:delayed}, left column). {(b,c)} Retrieved average values $\langle \alpha_R \rangle$ and $\langle \beta_R \rangle$ (only real parts are shown) from the resulting solution $A^{(R)}(t;\alpha,\beta)$ in the presence of noise. The averaging is performed with respect to 200 noise realizations for a noise amplitude $\delta\eta/\eta_0=0.05$. The input configurations are parameterized as  $\alpha=\cos(\vartheta/2)$ and $\beta=\sin(\vartheta/2)\,e^{i\varphi}$ with $\vartheta\in[0,\pi]$ and $\varphi\in[0,2\pi]$.  {(d)} Reconstructed Bloch sphere with spatial components $r_i=(\langle\alpha_R\rangle^*,\langle\beta_R\rangle^*)\cdot\sigma_i\cdot(\langle\alpha_R\rangle,\langle\beta_R\rangle)$ for $i=x,\,y,\,z$, evaluated from the retrieved averaged parameters taken from {(b,c)}, where $\sigma_i$ is the $i^{\mathrm{th}}$ Pauli matrix. The symbols in {(b-d)} emphasize the rebit encoding from {(a)}. The reference configurations, $\ket{0/1}$, are taken from the left column of Fig.~\ref{fig:sepzero:delayed}. {(e)} Maximum of the absolute errors, $\epsilon_\alpha=|\alpha-\langle \alpha_R \rangle|$ and $\epsilon_\beta=|\beta-\langle \beta_R \rangle|$, in retrieval of the input configurations for different noise amplitudes $\delta\eta$. Vertical dashed line shows the noise level of the calculations in (b-d).}
\label{fig-short-long-parameter-surface}
\vspace{-0.5cm}
\end{figure}

With these long coherence times we can now proceed to the main goal of storing coherent superpositions of the two spin configurations, $\ket{0/1}$. Those can be created by the corresponding superposition $\eta^{(W)}(t)=\alpha \cdot \eta_{\ket{0}}^{(W)}(t)+\beta \cdot \eta_{\ket{1}}^{(W)}(t)$ of the respective write pulses, and, ideally, the corresponding superposition of time-binned cavity responses would be observed under the application of the readout pulse $\eta^{(R)}(t)$. Since the cavity response is of the form $A^{(R)}(t;\alpha,\beta)\!=\!\alpha\! \cdot \! \tilde{A}^{(R)}_{\ket{0}}(t)+\beta\! \cdot \! \tilde{A}^{(R)}_{\ket{1}}(t)+\tilde{A}^{(R)}(t)$ where the two cavity responses $\tilde{A}^{(R)}_{\ket{0/1}}(t)$ only depend on the stored spin configurations $\ket{0/1}$, and $\tilde{A}^{(R)}(t)$ is the response induced by the readout pulse, the desired superposition of cavity outputs is obtained if the readout pulse satisfies $(\alpha+\beta)\eta^{(R)}(t)=\eta^{(R)}(t)$ (see Appendix \ref{App_B}). Together with the normalization $|\alpha|^2+|\beta|^2=1$ this implies that for the amplitudes $\alpha_x=1-x\pm i\sqrt{x(1-x)}$ and $\beta_x=x\mp i\sqrt{x(1-x)}$ with $x\in[0,1]$ the desired cavity response will be obtained. As a result, the proposed storage sequence does not only work for the two logical basis states $\ket{0/1}$, but, indeed for a one-dimensional set of coherent superpositions, such as for a rebit \cite{Brendel:1999,flopaper}.

Note that when being only interested in reading out the parameters $\alpha$ and $\beta$ (and not in further processing the resulting cavity response) one is not restricted by the above rebit parametrization, but has the full qubit parameter space at one's disposal. As we show in Appendix \ref{App_B}, $\alpha$ and $\beta$ can be unambiguously determined through the time-overlap integrals defined only in the readout section $[\tau_a,\tau_c]$ as ${\cal O}_{0/1}=\int_{\tau_a}^{\tau_c} dt\,A^{(R)}(t;\alpha,\beta)\cdot A^{(R)*}_{\ket{0/1}}(t)$, where $A^{(R)}_{\ket{0/1}}(t)=\tilde{A}^{(R)}_{\ket{0/1}}(t)+\tilde{A}^{(R)}(t)$.

In principle, this information retrieval is exact, but noise (which is not included in the previous theoretical modelling) affects the readout if it reaches values comparable to the cavity amplitudes. Therefore in the next line of our study we examine the robustness of our optimal control scheme against possible noise. For that purpose, we subject the previously established optimized pulses, $\eta^{(W)}_{\ket{0/1}}(t)$ and $\eta^{(R)}(t)$, to a small perturbation by adding Gaussian white noise as an additional driving term in our Volterra equations (see Appendix \ref{App_C}). We treat the problem numerically using well-established methods for integrating stochastic differential equations (see, e.g., \cite{Toral14}) and accumulate statistics by evaluating many trajectories for different noise realizations. We then average the resulting retrieved values with respect to noise realizations and calculate the absolute retrieval errors as the deviation from the input configuration, $\epsilon_\alpha=|\alpha-\langle \alpha_{{\cal R}} \rangle|$ and $\epsilon_\beta=|\beta-\langle \beta_{{\cal R}} \rangle|$. The typical results of our calculations are displayed in Fig.~\ref{fig-short-long-parameter-surface}. It turns out that $\epsilon_\alpha$ and $\epsilon_\beta$ scale approximately linearly with the noise amplitude and, e.g., the maximal absolute error of retrieval shown in Fig.~\ref{fig-short-long-parameter-surface} is at most $0.02$ for 200 noise realizations when taking the noise amplitude to be $5\%$ of the incoming amplitude of the write pulse. These results confirm the robustness of our approach with respect to possible noise in a real physical system.

\section{Conclusions and outlook}
In conclusion, we present here a very efficient optimization technique applicable to different experimental realizations based on an inhomogeneously broadened spin ensemble coupled to a single cavity mode. Generalizing this scheme to the full quantum mechanical level is the obvious next step to make our protocol an essential building block for the development of future optimal control schemes with the perspective of advancing the storage capabilities for quantum information. Given the extremely unfavorable scaling properties of composite quantum systems with particle number, any theoretical description of a quantum many-body system is an extremely challenging task. Since the identification of optimal control strategies is typically much harder than the mere description of a system's dynamics (the latter is naturally required for the former), optimal control is a viable option for rather small systems only. With our highly efficient semiclassical control technique for the non-Markovian dynamics of large hybrid quantum systems in the presence of inhomogeneous broadening, we demonstrate the capabilities and limitations of these systems for potential information storage.

\begin{acknowledgments}
We would like to thank Himadri Dhar and Matthias Zens for helpful discussions and acknowledge support by the Austrian Science Fund (FWF) through Project No.~F49-P10 (SFB NextLite). F.M. acknowledges support by the European Research Council (ERC) grant ``odycquent''.
\end{acknowledgments}

\appendix

\section{Volterra equation for the cavity amplitude}
\label{App_A}

Our starting point is the Hamiltonian (1) of the main article from which we derive the Heisenberg equations for the cavity and spin operators, $\dot a(t)=i [{\cal H},a(t)]-\kappa \cdot a(t)$, $\dot \sigma_k^{-}(t)=i[{\cal H},\dot \sigma_k^{-}(t)]-\gamma \cdot \sigma_k^{-}(t)$, respectively. Here $a$ stands for the cavity annihilation operator and $\sigma_k^{-}$ are standard downward Pauli operators associated with the $k$-th spin. $\kappa$ and $\gamma$ are the dissipative cavity and individual spin losses, respectively. (All notations are in tact with those introduced in the main article.)  During the derivations we use the following simplifications and approximations valid for various experimental realizations: (i) $k_B T\ll\hbar \omega_c$ (the energy of photons of the external bath, $k_B T$, is substantially smaller than that of cavity photons, $\hbar \omega_c$); (ii) the number of microwave photons in the cavity remains small as compared to the total number of spins participating in the coupling (limit of low input powers of an incoming signal), so that the Holstein-Primakoff-approximation, $\langle \sigma_k^{(z)} \rangle \approx -1$, always holds; (iii) the effective collective coupling strength of the spin ensemble, $\Omega^2=\sum_{k=1}^{N} g_k^{2}$ ($g_k$ stands for the coupling strength of the $k$th spin), satisfies the inequality $\Omega \ll \omega_c$, justifying the rotating-wave approximation; (iv) the spatial size of the spin ensemble is sufficiently smaller than the wavelength of a cavity mode. Having introduced all these assumptions, we derive the following system of coupled first-order linear ODEs for the cavity and spin amplitudes in the $\omega_p$-rotating frame
\begin{eqnarray}
\dot{A}(t)  &=&-[\kappa+i\Delta_c] A(t) + \sum_{k=1}^N g_k \, B_k(t) - \eta(t),
\label{Eq_a_Volt}
\\
\dot{B}_k(t)&=&-[\gamma+i\Delta_k] B_k(t) - g_k  \,A(t),
\label{Eq_bk_Volt}
\end{eqnarray}
where $A(t)\equiv \langle a(t)\rangle$ and $B_k(t)\equiv\langle\sigma_k^-(t)\rangle$.  $\Delta_c=\omega_c-\omega_p$ and $\Delta_k=\omega_k-\omega_p$ are the detunings with respect to the probe frequency $\omega_p$.

By formally integrating Eqs.~(\ref{Eq_bk_Volt}) with respect to time for the spin operators and inserting them into Eq.~(\ref{Eq_a_Volt}) for the cavity operator, we get
\begin{eqnarray}
\!\!\!\!\!\!\!\!\dot{A}(t)\!=\!-[\kappa+i\Delta_c] A(t) + \sum_{k=1}^{N} g_k  B_k(T_1)\ e^{-[\gamma+i\Delta_k] (t-T_1)}-\!\!\!\!\!\!\!\!\!\!\!\!\!\!\!
\nonumber
\\
\Omega^2\int\limits_{0}^{\infty} d\omega\, \rho(\omega) \int\limits_{T_1}^{t}d\tau\, A(\tau)
e^{-[\gamma+i\Delta_\omega](t-\tau)}-\eta(t),
\label{Eq_a_with_Bk0}
\end{eqnarray}
where $\Delta_\omega=\omega-\omega_p$, $B_k(T_1)$ are the initial spin amplitudes at $t=T_1$ and $\rho(\omega)=\sum_{k=1}^{N}  g_k^{2} \delta(\omega-\omega_k)/\Omega^2$ stands for the continuous spectral spin distribution. As in our previous studies \cite{Putz:2014aa,KPMS14}, we take into account the effect of an inhomogeneous broadening by modelling the spin density with a $q$-Gaussian shape, $\rho(\omega)=C\cdot\left[1-(1-q)(\omega-\omega_s)^2/\Delta^2\right]^{1/(1-q)}$, distributed around the mean frequency $\omega_s/2\pi=2.6915$\,GHz with the parameter $q=1.39$ and a full-width at half maximum (FWHM), $\gamma_q/2\pi=9.4$\,MHz, where $\gamma_q=2\Delta\sqrt{(2^q-2)/(2q-2)}$.

\begin{figure*}[t!]
\includegraphics[width=.75\textwidth]{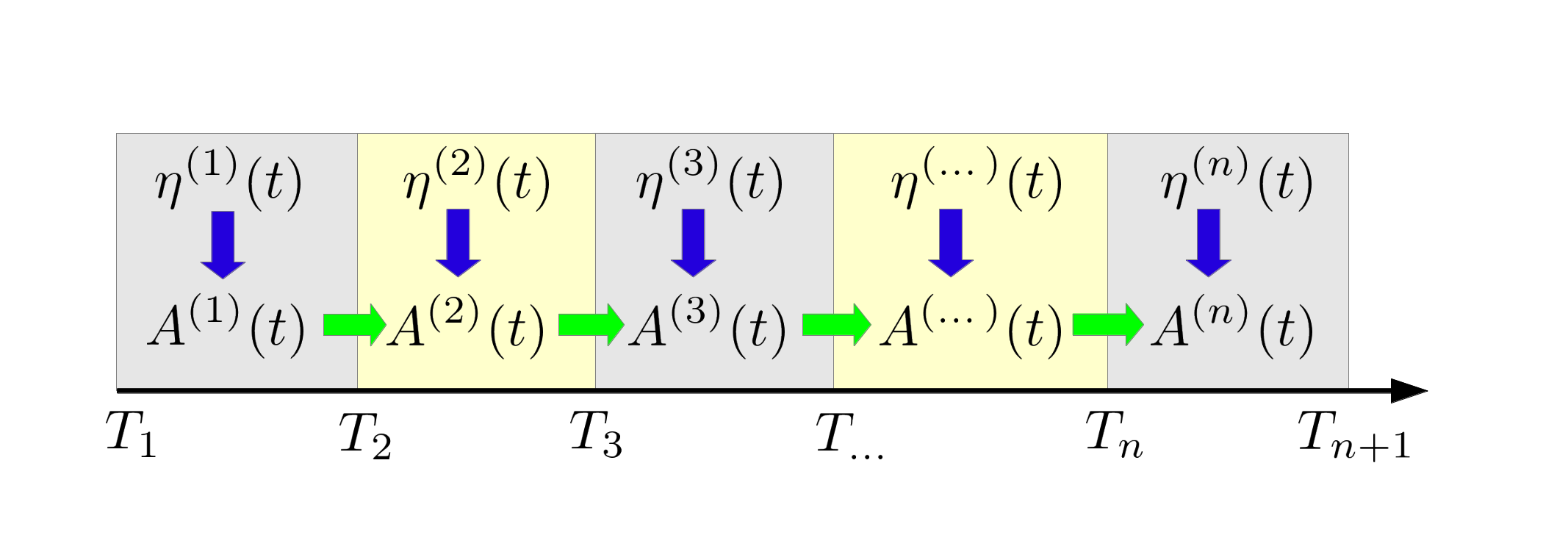}
%\vspace*{-0.75cm}
\caption{Schematics of the time-divisions of the cavity-amplitude $A^{(n)}(t)$. The input field $\eta^{(n)}(t)$ is applied to the system  in the time interval $[T_n,T_{n+1}]$ and drives the corresponding cavity-amplitude $A^{(n)}(t)$ (indicated by vertical arrows). The non-Markovian contributions from previous time intervals $[T_{n-1},T_n]$ are indicated by horizontal  arrows.}
\label{fig:notation:timeinterval}
\end{figure*}

Next we formally integrate Eq.~(\ref{Eq_a_with_Bk0}) in time and simplify the resulting double integral on the right-hand side by partial integration. We also consider the case when the cavity is initially empty, $A(T_1)=0$, and all spins are in the ground state, $B_k(T_1)=0$. To speed up our numerical calculations and to separate different time sections from each other (see the main text and Appendix \ref{App_B} for details), we divide the whole time integration into successive subintervals, $T_{ n}\leq t \leq T_{n+1}$, with $n=1,2,...$ (see Fig.~\ref{fig:notation:timeinterval}). This allows us to derive the recurrence relation for the cavity amplitude for the $n$-th time interval, $A^{(n)}(t)$, which depends on all previous events at $t<T_n$. Finally, we end up with the following expression for $A^{(n)}(t)$
\begin{eqnarray}
\label{Eq_A_subs_main}
A^{(n)}(t)=\int\limits_{T_n}^t d\tau \, {\cal K}(t-\tau) A^{(n)}(\tau)+{\cal D}^{(n)}(t)+{\cal F}^{(n)}(t),\,\,\,\,\,\,\,\,\,\,\,\,
\end{eqnarray}
where the non-Markovian feedback within the $n$-th time interval is provided by the kernel function ${\cal K}(t-\tau)$
\begin{eqnarray}
\label{eq:eqm:A:Volterra:kernel}
\!\!{\cal K}(t-\tau)\!=\!\Omega^2\int\limits_0^{\infty}\! d\omega\rho(\omega)
\dfrac{
e^{-[\gamma+i\Delta_{\omega}](t-\tau)}
\!
-
\!
e^{[\kappa+i\Delta_c](t-\tau)}
}
{\left[\gamma+i\,\Delta_\omega\right]\!-\!\left[\kappa+i\,\Delta_c\right]}.\,\,\,\,\,\,\,\,\,
\end{eqnarray}
The driving term ${\cal D}^{(n)}(t)$ in Eq.~(\ref{Eq_A_subs_main}),
\begin{eqnarray}
{\cal D}^{(n)}(t) = \, -\int\limits_{T_n}^t d\tau\, \eta^{(n)}(\tau)\,e^{-[\kappa+i\Delta_c](t-\tau)},
\label{eq:eqm:A:Volterra:driving}
\end{eqnarray}
includes an arbitrarily shaped, weak incoming-pulse $\eta^{(n)}(t)$, defined in the time interval
$[T_n,T_{n+1}]$. The memory contributions from all previous time intervals for $t<T_n$ are given both through the amplitude $A^{(n-1)}(T_n)$ and through the memory integral ${\cal I}^{(n)}(\omega)$, which are contained in the function
\begin{widetext}
\begin{multline}
{\cal F}^{(n)}(t) =
      \left\{\!\!\!\!\!\!\phantom{\int\limits_0^{\infty}}
      A^{(n-1)}(T_n)\,e^{-[\kappa+i\Delta_c] (t-T_n)}
      +\,
      \Omega^2\int\limits_0^{\infty} d\omega\, \rho(\omega)
      \dfrac{
        e^{-[\gamma+i\Delta(\omega)](t-T_n)}
        -
        e^{-[\kappa+i\Delta_c](t-T_n)}
      }
      {
        \left[\gamma+i\,\Delta_\omega\right]-\left[\kappa+i\,\Delta_c\right]
      }
      \cdot  {\cal I}^{(n)}(\omega)
\right\}, \,\,\,\,\,\,\,\,\,\,\,\,\,\,\,\,\,\,\,\,\,\,\,\,\,\,\,\,\,\,
\label{eq:eqm:A:Volterra:memory}
\end{multline}
where
\begin{equation}
{\cal I}^{(n)}(\omega) =  {\cal I}^{(n-1)}(\omega)\,e^{-[\gamma+i\Delta_\omega](T_n-T_{n-1})}
                          \, + \int\limits_{T_{n-1}}^{T_n} d\tau\,
                          A^{(n-1)}(\tau)\,e^{-[\gamma+i\Delta_\omega](T_n-\tau)},
\label{eq:eqm:A:Volterra:memory:recursive}
\end{equation}
\end{widetext}
In accordance with the initial conditions introduced above at $t=T_1$, $A^{(0)}(T_1)=0$ and ${\cal I}^{(1)}(\omega)=0$, so that ${\cal F}^{(1)}(t)$ vanishes in the first time interval, ${\cal F}^{(1)}(t)=0$ ($T_1 \le t \le T_2$).

\begin{figure*}
\includegraphics*[width=0.85\textwidth]{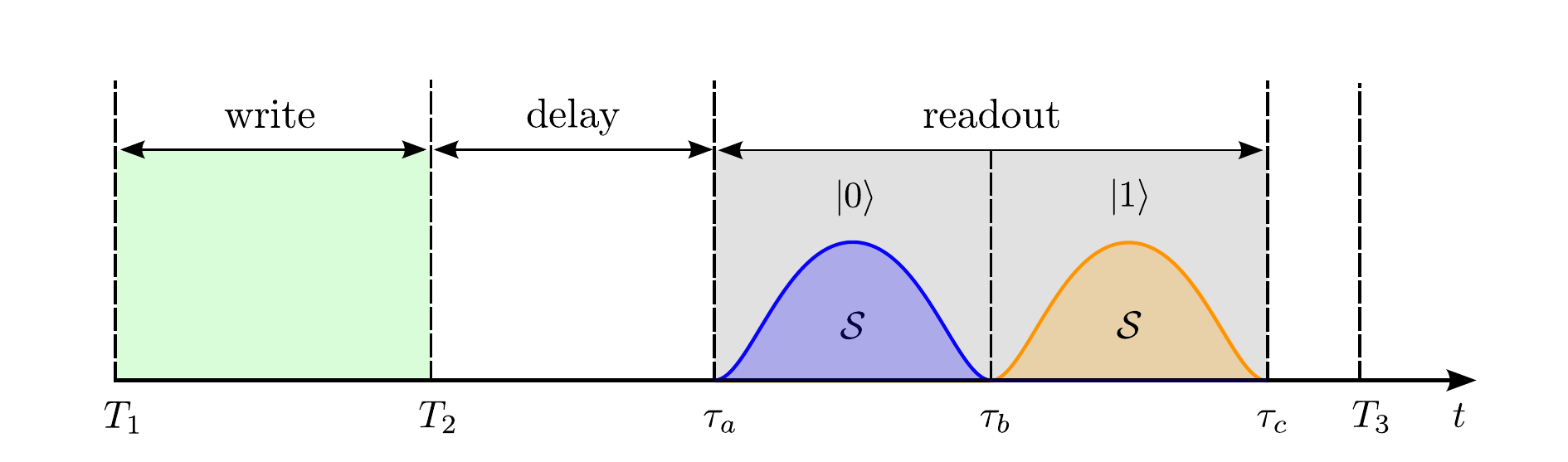}
\caption{Time divisions for the optimization scheme of the cavity responses $A_{\ket{0}}^{(R)}(t)$ and $A_{\ket{1}}^{(R)}(t)$. The {\it write section}, $[T_1,T_2]$, is followed by the variable {\it delay section}, $[T_2,\tau_a]$, and the {\it readout section} $[\tau_a, \tau_c]$. $A_{\ket{0}}^{(R)}(t)$ and $A_{\ket{1}}^{(R)}(t)$ reside the first, $[\tau_a, \tau_b]$, and the second half, $[\tau_b, \tau_c]$, of the {\it readout section}, respectively.}
\label{fig-timedivisions-results}
\end{figure*}
\section{Optimal control based on the Volterra equation}
\label{App_B}

In the main text of the manuscript, we split our time interval into two parts, a {\it write} and {\it readout section}, with a variable {\it delay section} in between. In the {\it write section}, two independent optimized write pulses $\eta^{(W)}(t)$ prepare two different configurations of the spin ensemble, which are referred to as logical states $\ket{0}$ and $\ket{1}$ of the spin ensemble. It is followed by the {\it delay section} characterized by almost completely suppressed cavity responses, and finally by the {\it readout section} where two logical states of the spin ensemble are retrieved and mapped on two mutually orthogonal states of the cavity field by means of the readout pulse $\eta^{(R)}(t)$ (see Fig.~\ref{fig-timedivisions-results}). Note that the optimized readout pulse is generic being the same for both $\ket{0}$ and $\ket{1}$ states. For the sake of simplicity we do not explicitly specify the delay pulse but impose on the readout pulse $\eta^{(R)}(t)$ a constraint such that the cavity responses are maximally suppressed in the {\it delay section}  $[T_2,\tau_a]$, see Fig.~\ref{fig-timedivisions-results}. Thus, the write and readout pulses are defined within the time intervals, $[T_1,T_2]$ and $[T_2,T_3]$, respectively, in terms of the notations introduced in the Appendix \ref{App_A} and the {\it delay section} is formally absorbed into the {\it readout section}.

We then expand $\eta^{(W)}(t)$ and $\eta^{(R)}(t)$ in terms of sine functions
\begin{eqnarray}
\label{eta_W}
\eta^{(W)}(t)&=&\sum_{k=1}^{N_1} \xi_k\cdot \sin(k\,\omega_f (t-T_1)),
\\
\label{eta_R}
\eta^{(R)}(t)&=&\sum_{l=1}^{N_2} \zeta_l\cdot \sin(l\,\omega_f (t-T_2)),
\end{eqnarray}
where $\xi_k$ and $\zeta_l$ are the expansion coefficients and $\omega_f$ is the fundamental frequency. The linear property of the Volterra equation (\ref{Eq_A_subs_main}) allows us to expand the cavity amplitude in the {\it write section}, $A^{(W)}(t)$, in a series of time-dependent functions with the same expansion coefficients $\xi_k$ as in Eq.~(\ref{eta_W}),
\begin{eqnarray}
\label{AW_Ansatz}
A^{(W)}(t)=\sum_{k=1}^{N_1}  \xi_k \cdot a_k^{(W)}(t).
\end{eqnarray}
Here $a_k^{(W)}(t)$ are solutions of the following Volterra equations
\begin{eqnarray}
a_k^{(W)}(t)=\int\limits_{T_1}^t d\tau\,{\cal K}(t-\tau)\,a_k^{(W)}(\tau)-\nonumber\\
\int\limits_{T_1}^t d\tau\, \sin(k\,\omega_f (\tau-T_1))\,e^{-[\kappa+i \Delta_c](t-\tau)},
\label{ak_W}
\end{eqnarray}
where the kernel function ${\cal K}(t-\tau)$ is given by Eq.~(\ref{eq:eqm:A:Volterra:kernel}).

The solution in the {\it readout section}, $A^{(R)}(t)$, in turn, consists of two contributions
\begin{eqnarray}
\label{AR_Ansatz}
A^{(R)}(t)=\sum_{l=1}^{N_2} \zeta_l \, a_l^{(R)}(t)+\sum_{k=1}^{N_1} \xi_k \, \psi_k^{(R)}(t)  \textnormal{.}
\end{eqnarray}
Similar to the ansatz for the {\it write section}, the first term in Eq.~(\ref{AR_Ansatz})
also contains the same expansion coefficients $\zeta_l$ as the corresponding driving signal in the {\it readout section} (see Eq.~(\ref{eta_R})) with the time-dependent functions, $a_l^{(R)}(t)$, obeying the following Volterra equations ($T_2\le t \le T_3$)
\begin{eqnarray}
a_l^{(R)}(t)=\int\limits_{T_2}^t d\tau\,{\cal K}(t-\tau)\,a_l^{(R)}(\tau)-
\nonumber\\
\int\limits_{T_2}^t d\tau\,
\sin(l\, \omega_f (\tau-T_2))\,e^{-[\kappa+i \Delta_c](t-\tau)}.
\label{al_Harm}
\end{eqnarray}
Additionally, the second term in Eq.~(\ref{AR_Ansatz}) describes the non-Markovian memory and appears in the {\it readout  section} due to the energy stored both in the cavity and spin ensemble during the time interval $T_1\le t \le T_2$ ({\it write section}). Therefore, it depends only on the coefficients $\xi_k$ of the {\it write pulse} (\ref{eta_W}) and the time-dependent functions $\psi_k^{(R)}(t)$, which can be found by substituting the expressions (\ref{AW_Ansatz},\ref{AR_Ansatz}) into Eqs.~(\ref{Eq_A_subs_main}-\ref{eq:eqm:A:Volterra:memory:recursive}) for $n=2$.  It can be shown that these functions satisfy the following Volterra equations
\begin{eqnarray}
\label{eq:eqm:A:Volterra:region:1:harmonic:k}
\psi_k^{(R)}(t)=\int\limits_{T_2}^t d\tau\,{\cal K}(t-\tau)\,\psi_k^{(R)}(\tau)+f_k^{(R)}(t),
\end{eqnarray}
with the feedback from the previous {\it write section} defined by
\begin{widetext}
\begin{eqnarray}
\label{fk_R}
f_k^{(R)}(t)\!=\!a_k^{(W)}(T_2)\,e^{-[\kappa+i\Delta_c](t-T_2)}+
      \Omega^2\int\limits_0^{\infty} d\omega\, \rho(\omega)
      \dfrac{e^{-[\gamma+i\Delta_\omega](t-T_2)}-e^{-[\kappa+i\Delta_c](t-T_2)}}
      {\left[\gamma+i\,\Delta_\omega\right]-\left[\kappa+i\,\Delta_c\right]}\!\cdot\!\!\!
\int\limits_{T_1}^{T_2} d\tau\, a_k^{(W)}(\tau)\,e^{-[\gamma+i\Delta_\omega](T_2-\tau)}.\,\,\,\,\,\,\,\,\,\,
\end{eqnarray}
\end{widetext}
Note that $a_k^{(W)}(t)$ in Eq.~(\ref{fk_R}) are defined in the {\it write section} only and are known solutions of Eq.~(\ref{ak_W}).

In the main text of our manuscript we use two different pulses $\eta^{(W)}_{\ket{0}}(t)=\sum_{k=1}^{N_1}  \xi_k^{\ket{0}}\cdot \sin(k\,\omega_f (t-T_1))$ and $\eta^{(W)}_{\ket{1}}(t)=\sum_{k=1}^{N_1}  \xi_k^{\ket{1}}\cdot \sin(k\,\omega_f (t-T_1))$ in the {\it write section}, which are characterized by two sets of expansion coefficients from  Eq.~(\ref{eta_W}). As a result, the cavity amplitudes in the {\it write section} are also represented by these sets of expansion coefficients and are given by Eq.~(\ref{AW_Ansatz}), namely
\begin{eqnarray}
A^{(W)}_{\ket{0}}(t)=\sum_{k=1}^{N_1}  \xi_k^{\ket{0}} \cdot a_k^{(W)}(t),
\nonumber\\
A^{(W)}_{\ket{1}}(t)=\sum_{k=1}^{N_1}  \xi_k^{\ket{1}} \cdot a_k^{(W)}(t).
\label{AW_Ansatz_0_1}
\end{eqnarray}
Note that by injecting these pulses into the cavity, we create two independent configurations (denoted as $\ket{0}$ and
$\ket{1}$) of the spin-cavity system at the beginning of the readout interval, $t=T_2$.

Next, we perform a readout by applying a single optimized readout pulse (\ref{eta_R}), which is the same for the states $\ket{0}$ and $\ket{1}$. The cavity amplitudes in the {\it readout section}, in turn, are governed by Eq.~(\ref{AR_Ansatz}) as
\begin{eqnarray}
A^{(R)}_{\ket{0}}(t)=\underbrace{\sum_{l=1}^{N_2} \zeta_l \, a_l^{(R)}(t)}_{\tilde{A}^{(R)}(t)}+\underbrace{\sum_{k=1}^{N_1} \xi_k^{\ket{0}} \, \psi_k^{(R)}(t)}_{\tilde{A}^{(R)}_{\ket{0}}(t)},\,\,\,\,
\nonumber\\
A^{(R)}_{\ket{1}}(t)=\underbrace{\sum_{l=1}^{N_2} \zeta_l \, a_l^{(R)}(t)}_{\tilde{A}^{(R)}(t)}+\underbrace{\sum_{k=1}^{N_1} \xi_k^{\ket{1}} \, \psi_k^{(R)}(t)}_{\tilde{A}^{(R)}_{\ket{1}}(t)},\,\,\,\,\,\,\,\,\,\,
\label{AR_Ansatz_0_1}
\end{eqnarray}
where $\tilde{A}^{(R)}(t)$ describes the contribution from the {\it readout} pulse only which is the same for both cavity responses and the two other terms, $\tilde{A}^{(R)}_{\ket{i}}(t)$ ($i=0,1$), explicitly depend on the states $\ket{0}$ and
$\ket{1}$ created in the {\it write section}.

Thus, the cavity amplitude is determined at every moment of time by Eqs.~(\ref{AW_Ansatz}-\ref{fk_R}) (and, as a consequence, all spin configurations), if all expansion coefficients, $\xi_k^{\ket{0}}\!$, $\xi_k^{\ket{1}}\!$, and $\zeta_l$ are provided.

As a next step, we develop an optimization scheme aiming at achieving two well-resolved cavity responses in the {\it readout section}, $A^{(R)}_{\ket{0}}(t)$ and $A^{(R)}_{\ket{1}}(t)$, as is sketched in Fig.~\ref{fig-timedivisions-results}. (The results of numerical calculations are presented in Fig.~2 of the main paper.)  For this purpose we use the standard method of Lagrange multipliers by introducing the functional ${\cal F}(\xi_k^{\ket{0}}\!, \xi_k^{\ket{1}}\!, \zeta_l)$ subject to several constraints listed below, and search for its minima with respect to the expansion coefficients of all three pulses. Namely, we write the following expression for the functional
\begin{eqnarray}
\nonumber
&&{\cal F}(\xi_k^{\ket{0}}\!, \xi_k^{\ket{1}}\!, \zeta_l)\!=\!
\int_{\tau_b}^{\tau_c} dt\,|A^{(R)}_{\ket{0}}(t)|^2\!+\! \\\nonumber
&&\int_{\tau_a}^{\tau_b} dt\,|A^{(R)}_{\ket{1}}(t)|^2\!+\!
\left|\int_{\tau_a}^{\tau_c} dt\,A^{(R) \star}_{\ket{0}}(t)A^{(R)}_{\ket{1}}(t)\right|\!-\\\nonumber
&&\lambda_{delay}^{\ket{0}}\!\cdot \int_{T_2}^{\tau_a} dt\, |A^{(R)}_{\ket{0}}(t)|^2\!-\!
\\\nonumber
&&\lambda_{delay}^{\ket{1}}\!\cdot \int_{T_2}^{\tau_a} dt\, |A^{(R)}_{\ket{1}}(t)|^2\!-
\lambda_{T}^{\ket{0}}\cdot |A^{(R)}_{\ket{0}}(\tau_a)|^2\!-\\\nonumber
&&\!\lambda_{T}^{\ket{1}}\cdot|A^{(R)}_{\ket{1}}(\tau_a)|^2\!-\!
\lambda_{\Delta T}^{\ket{0}}\!\cdot\!\left( \int_{\tau_a}^{\tau_b} dt\, |A^{(R)}_{\ket{0}}(t)|^2-{\cal S}\right)\!-
\\
&&\!\lambda_{\Delta T}^{\ket{1}}\!\cdot\!\left( \int_{\tau_b}^{\tau_c} dt\, |A^{(R)}_{\ket{1}}(t)|^2-{\cal S}\right)\!-\nonumber\\
&&\!\lambda_{P}^{\ket{0}}\!\cdot\! \left( \sum_k |\xi_k^{\ket{0}}|^2-{\cal P}\right)
\!-\!\lambda_{P}^{\ket{1}}\!\cdot\! \left( \sum_k |\xi_k^{\ket{1}}|^2-{\cal P}\right)\!\!,\,\,\,\,\,\,\,\,\,\,\,
\label{F_Lagr_mult}
\end{eqnarray}
where $\lambda$-s are the Lagrange multipliers. The first three terms in Eq.~(\ref{F_Lagr_mult}) are the functions to be minimized which ensure that the overlap between the time-binned states in the {\it readout section} is negligibly small. The rest of the terms are constraints which additionally guarantee the following conditions to be simultaneously fulfilled: (i) the cavity responses within the {\it delay section} are maximally suppressed; (ii) the cavity at the beginning of the {\it readout section} is almost empty for both states; (iii) the integral taken with respect to the time-binned cavity amplitudes squared within the {\it readout section} has the same value ${\cal S}$; (iv) a net power ${\cal P}$ of the {\it write} pulses per fundamental period $2\pi/\omega_f$ is the same.

In our numerical calculations we used the sequential Least Squares Programming (SLSQP) minimization method \cite{Kraft88} embedded in the internal python library {\it scipy.optimize} to find the minima of the functional ${\cal F}(\xi_k^{\ket{0}}\!, \xi_k^{\ket{1}}\!, \zeta_l)$.

In the main text we create an arbitrary superposition of write pulses (each of which separately prepares the logical state $\ket{0}$ or $\ket{1}$) by applying the superimposed write pulse
\begin{eqnarray}
\eta^{(W)}(t)=\alpha\cdot\eta_{\ket{0}}^{(W)}(t)+\beta \cdot\eta_{\ket{1}}^{(W)}(t)
\end{eqnarray}
aiming to extract the encoded information (given by complex numbers $\alpha$ and $\beta$) from the solution for the cavity amplitude in the {\it readout section} designated in Fig.~\ref{fig-timedivisions-results}. (Note that the reading pulse $\eta^{(R)}(t)$ is always kept the same.) The solution in the {\it readout section} can be written as
\begin{eqnarray}
A^{(R)}(t;\alpha,\beta)\!=\!\alpha\! \cdot \! \tilde{A}^{(R)}_{\ket{0}}(t)+\beta\! \cdot \! \tilde{A}^{(R)}_{\ket{1}}(t)+\tilde{A}^{(R)}(t),\,
\label{A_gen}
\end{eqnarray}
where all three previously established well-known amplitudes $\tilde{A}^{(R)}_{\ket{0}}(t)$, $\tilde{A}^{(R)}_{\ket{1}}(t)$ and $\tilde{A}^{(R)}(t)$ are introduced in Eq.~(\ref{AR_Ansatz_0_1}). We then project our resulting solution (\ref{A_gen}) onto the functions $A^{(R)}_{\ket{0}}(t)$ and $A^{(R)}_{\ket{1}}(t)$ from Eq.~(\ref{AR_Ansatz_0_1}), namely, we write
\begin{eqnarray}
{\cal O}_{i}=\int\limits_{\tau_a}^{\tau_c}dt\,A^{(R)}(t;\alpha,\beta)\cdot A^{(R)*}_{\ket{i}}(t)=
\nonumber\\
\alpha\cdot{\cal F}_{i,0} + \beta\cdot{\cal F}_{i,1} + {\cal F}_{i,R},
\label{eq:super:reg:2:overlap}
\end{eqnarray}
where the overlap integrals $ {\cal F}_{i,q}=\int\limits_{\tau_a}^{\tau_c} dt\,\tilde{A}^{(R)}_{\ket{q}}(t)\cdot A^{(R)*}_{\ket{i}}(t) $ with $i,q=0,1$ and ${\cal F}_{i,R}=\int\limits_{\tau_a}^{\tau_c} dt\,\tilde{A}^{(R)}(t)\cdot A^{(R)*}_{\ket{i}}(t)
$. Since ${\cal F}_{i,q}$ and ${\cal F}_{i,R}$ are known we finally end up with the following set of two algebraic equations
\begin{eqnarray}
\label{Eq_O1}
{\cal O}_{0}&=&\alpha\cdot{\cal F}_{0,0} + \beta\cdot{\cal F}_{0,1} + {\cal F}_{0,{R}},\\
{\cal O}_{1}&=&\alpha\cdot{\cal F}_{1,0} + \beta\cdot{\cal F}_{1,1} + {\cal F}_{1,{R}},
\label{Eq_O2}
\end{eqnarray}
from which the retrieved values $\alpha_R$ and $\beta_R$ can be evaluated.

\section{Retrieval of encoded parameters in the presence of noise}
\label{App_C}
%\label{Section_Volt_ampl}
%

Here we study the influence of noise on the quality of our optimization scheme presented in the main article and introduced in the Appendix \ref{App_B}. For that purpose, we subject the previously established optimal driving amplitudes, $\eta^{(W)}_{\ket{0}}(t)$, $\eta^{(W)}_{\ket{1}}(t)$ and $\eta^{(R)}(t)$ (see Appendix \ref{App_B}), to a small perturbation represented by the driving term, $\delta \eta_{noise}(t)=\delta \eta\cdot \upsilon(t)$, where $\delta \eta$ is the amplitude of perturbation and $\upsilon(t)$ stands for a Gaussian white noise of mean and correlations given by, respectively, $\langle \upsilon(t) \rangle$=0 and $\langle \upsilon(t')  \upsilon(t) \rangle=\delta(t-t')$. We then numerically integrate the Volterra equation (\ref{Eq_a_with_Bk0}) from Appendix \ref{App_A} with respect to time by adding the perturbation $\delta \eta_{noise}(t)$ to the corresponding deterministic optimal driving amplitudes $\eta(t)$, which in our specific case are represented by the known writing and readout amplitudes, $\eta^{(W)}_{\ket{0}}(t)$, $\eta^{(W)}_{\ket{1}}(t)$ and $\eta^{(R)}(t)$. We treat the problem numerically using well-established numerical methods for integrating stochastic differential equations (see e.g. \cite{Toral14}). In a nutshell, the stochastic contribution to the cavity amplitude is taken into account after each time step of numerical integration in the following way: $A(t_{m+1}) \rightarrow  A(t_{m+1}) + \sqrt{dt}\cdot \delta\eta_{noise}(t_m)$, where $A(t_{m+1})$ after the arrow corresponds to the deterministic part of the cavity amplitude at $t=t_{m+1}$ obtained using the standard Runge-Kutta method and $\delta \eta_{noise}(t_m)$ is the stochastic drive taken from the previous time step. We then accumulate statistics by integrating many trajectories for different noise realizations. Next, we extract the encoded parameters $\alpha_{{\cal R}}$ and $\beta_{{\cal R}}$ in the presence of noise replacing the overlap integrals in Eqs.~(\ref{Eq_O1}, \ref{Eq_O2}) for the case without noise by the corresponding overlap integrals evaluated for different noise realizations. The result of calculations for the average retrieval values of $\langle \alpha_{{\cal R}} \rangle$ and $\langle \beta_{{\cal R}} \rangle$ and their absolute errors, $\epsilon_\alpha=|\alpha-\langle \alpha_{{\cal R}} \rangle|$ and $\epsilon_\beta=|\beta-\langle \beta_{{\cal R}} \rangle|$, with respect to the encoded values are depicted in Fig.~3 of the main paper.

\section{Numerical values for the optimized readout pulse coefficients}
\label{App_D}

Here we present numerical values of the coefficients $\xi_k^{|0\rangle}$, $\xi_k^{|1\rangle}$ and $\zeta_l$ of the optimal readout pulses $\eta^{(W)}_{|0\rangle}(t)$, $\eta^{(W)}_{|1\rangle}(t)$ and $\eta^{(R)}(t)$ defined by Eqs.~(\ref{eta_W}-\ref{eta_R}), which are presented in the main text. We take the amplitude of the write pulses such that the net power injected into the cavity, ${\cal P}^{(W)}_{|i\rangle}=\frac{1}{T_f}\int\limits_0^{T_f}dt\,|\eta^{(W)}_{|i\rangle}(t)|^2=\kappa^2$, with $i=0,1$, such that it corresponds to the power provided by a coherent driving signal with the amplitude equal to the cavity decay rate, $\eta=\kappa$.  Specifically, using the expansion (\ref{eta_W}) for the write pulses $\eta^{(W)}(t)$, we obtain the following expression for the power of the write pulses per fundamental period $T_f$:
\begin{eqnarray}
{\cal P}^{(W)}_{|i\rangle}=\eta^{(W)2}\cdot\frac{1}{2}\sum_{k=1}^{N_1} \left |\xi_k^{|i\rangle}/\eta^{(W)} \right |^2 = \kappa^2,
\end{eqnarray}
where $\eta^{(W)}=\kappa$ and $1/2\cdot\sum_{k=1}^{N_1} | \xi_k^{|i\rangle}/\kappa |^2=1$ due to the constraint imposed on the expansion coefficients. On the other hand the power of the readout pulse is substantially smaller than that of the write pulses and for the case without hole burning (see left column of Fig.~2 in the main text) we obtain
\begin{eqnarray}
{\cal P}^{(R)}=\eta^{(R)2}\cdot\frac{1}{2}\sum_{k=1}^{N_2} \left | \zeta_l/\eta^{(R)} \right |^2 = 0.068\cdot \kappa^2,
\end{eqnarray}
where $\eta^{(R)}=0.26\cdot \kappa$ and again we use as the constraint $1/2\cdot \sum_{l=1}^{N_2} |\zeta_l/\eta^{(R)}|^2=1$.

The coefficients for all optimal readout pulses shown in the left column of Fig.~2 (main text of the paper) are listed in Table.~\ref{tbl-pulse-coefficients}. For the sake of convenience the coefficients of the write and readout pulses are normalized to  $\eta^{(W)}$ and $\eta^{(R)}$, respectively. We use $N_1=5$ coefficients for the write pulse and $N_2=10$ for the readout pulse (notation is consistent with that used in Appendices \ref{App_A}, \ref{App_B}). The fundamental frequency for the write pulses is given by $\omega_f=\pi/(T_2-T_1)=\Omega_R$ and for the readout pulse we use $\omega_f=\pi/(T_3-T_2)=\Omega_R/2$. Here the Rabi-frequency $\Omega_R=2\pi\cdot 13.62\,$MHz and the time divisions shown in Fig.~\ref{fig-timedivisions-results} are $T_1=0$, $T_2=36.72\,$ns and $T_3=110.15\,$ns. The readout section, $[\tau_a,\tau_c]$, coincides with the whole readout interval, $[T_2,T_3]$.
\begin{widetext}
\begin{table*}[t]%
  \centering
	\begin{tabular}{r||r||r|r}
\thead{$\xi_{k=1\dotsc 5}^{|0\rangle}$} & \thead{$\xi_{k=1\dotsc 5}^{|1\rangle}$} & \thead{$\zeta_{l=1\dotsc 5}^{\phantom{|\rangle}}$} & \thead{$\zeta_{l=6\dotsc 10}^{\phantom{|\rangle}}$}\\\hline\hline
$ 0.434 +0.103~i$ & $-0.043 -0.013~i$ & $-1.003 -0.250~i$ & $ 0.229 +0.054~i$\\
$ 0.303 +0.067~i$ & $-0.231 -0.055~i$ & $ 0.820 +0.195~i$ & $ 0.037 +0.007~i$\\
$ 1.060 +0.259~i$ & $-1.127 -0.273~i$ & $-0.017 -0.007~i$ & $-0.096 -0.025~i$\\
$-0.152 -0.023~i$ & $ 0.200 +0.044~i$ & $-0.213 -0.054~i$ & $-0.174 -0.043~i$\\
$ 0.682 +0.161~i$ & $-0.723 -0.175~i$ & $-0.243 -0.061~i$ & $ 0.105 +0.024~i$\\
  \end{tabular}
\caption{
				 Normalized expansion coefficients
				 $\xi_{k=1,\dotsc,5}^{|i\rangle}$ (for $i=0,1$) and $\zeta_{k=1,\dotsc,10}$
				 defined by Eqs.~(\ref{eta_W}-\ref{eta_R}),
				 which correspond to the optimal readout pulses $\eta^{(W)}_{|0\rangle}(t)$, $\eta^{(W)}_{|1\rangle}(t)$ and $\eta^{(R)}(t)$
				 depicted in the left column of Fig.~2 of the main text. 				 The coefficients for the write pulses are normalized to
				 $\eta^{(W)}=\kappa$, and for the readout pulse to $\eta^{(R)} = 0.26\cdot\kappa$.
         }
\label{tbl-pulse-coefficients}
\end{table*}
\end{widetext}
For the case with hole-burning, depicted on the right column of Fig.~2, we use $N_1=4$ and $N_2=60$. All coefficients are summarized in Table~\ref{tbl-pulse-coefficients-holes}. Here we choose the fundamental frequency for the write pulses as $\omega_f=\pi/(T_2-T_1)=\Omega_R/2$, whereas  $\omega_f=\pi/(T_3-T_2)=\Omega_R/30$ for the readout pulse. The time-divisions are $T_1=0$, $T_2=73.4$~ns and $T_3=1174.9$~ns and the Rabi-frequency $\Omega_R=2\pi\cdot 13.62$~MHz. The readout section defined by $\tau_a=1114.3$~ns and $\tau_c=1153.6$~ns, is delayed by approximately $1$~$\mu$s with respect to the write section, $[T_1,T_2]$. The power ratio of the readout pulse to the write pulse turns out to be ${\cal P}^{(R)} / {\cal P}^{(W)} = 0.013$.
\begin{widetext}
\begin{table*}[t] %
  %\centering
	\begin{tabular}{r||r||r|r|r|r}
\thead{$\xi_{k=1\dotsc 4}^{|0\rangle}$} & \thead{$\xi_{k=1\dotsc 4}^{|1\rangle}$} & \thead{$\zeta_{l=1\dotsc 15}^{\phantom{|\rangle}}$} & \thead{$\zeta_{l=16\dotsc 30}^{\phantom{|\rangle}}$} & \thead{$\zeta_{l=31\dotsc 45}^{\phantom{|\rangle}}$} & \thead{$\zeta_{l=46\dotsc 60}^{\phantom{|\rangle}}$}\\\hline\hline
$-0.227 +0.108~i$ & $ 1.252 -0.161~i$ & $ 0.066 -0.028~i$ & $-0.239 -0.024~i$ & $ 0.054 -0.032~i$ & $-0.036 -0.010~i$\\
$ 0.017 +0.046~i$ & $ 0.074 +0.050~i$ & $-0.121 -0.022~i$ & $ 0.228 -0.088~i$ & $-0.041 -0.014~i$ & $ 0.059 -0.021~i$\\
$ 1.014 -0.161~i$ & $-0.243 +0.083~i$ & $ 0.190 -0.128~i$ & $-0.170 -0.013~i$ & $ 0.044 -0.029~i$ & $-0.069 -0.011~i$\\
$ 0.938 -0.032~i$ & $ 0.574 +0.040~i$ & $-0.230 +0.010~i$ & $ 0.161 -0.080~i$ & $-0.027 -0.014~i$ & $ 0.090 -0.026~i$\\
$               $ & $               $ & $ 0.292 -0.151~i$ & $-0.109 -0.019~i$ & $ 0.027 -0.025~i$ & $-0.096 -0.007~i$\\
$               $ & $               $ & $-0.313 +0.033~i$ & $ 0.107 -0.059~i$ & $-0.007 -0.015~i$ & $ 0.112 -0.040~i$\\
$               $ & $               $ & $ 0.365 -0.155~i$ & $-0.066 -0.020~i$ & $ 0.006 -0.020~i$ & $-0.112 +0.016~i$\\
$               $ & $               $ & $-0.363 +0.028~i$ & $ 0.073 -0.051~i$ & $ 0.012 -0.020~i$ & $ 0.122 -0.051~i$\\
$               $ & $               $ & $ 0.398 -0.160~i$ & $-0.042 -0.021~i$ & $-0.009 -0.017~i$ & $-0.118 +0.004~i$\\
$               $ & $               $ & $-0.377 -0.001~i$ & $ 0.058 -0.044~i$ & $ 0.021 -0.022~i$ & $ 0.122 -0.033~i$\\
$               $ & $               $ & $ 0.395 -0.136~i$ & $-0.032 -0.021~i$ & $-0.012 -0.009~i$ & $-0.115 +0.011~i$\\
$               $ & $               $ & $-0.355 -0.009~i$ & $ 0.025 -0.040~i$ & $ 0.016 -0.021~i$ & $ 0.117 -0.031~i$\\
$               $ & $               $ & $ 0.358 -0.097~i$ & $-0.085 -0.017~i$ & $ 0.000 -0.014~i$ & $-0.108 +0.001~i$\\
$               $ & $               $ & $-0.305 -0.001~i$ & $ 0.049 -0.033~i$ & $-0.004 -0.018~i$ & $ 0.107 -0.032~i$\\
$               $ & $               $ & $ 0.298 -0.093~i$ & $-0.047 -0.015~i$ & $ 0.026 -0.023~i$ & $-0.095 +0.020~i$\\
  \end{tabular}
\caption{
				 Normalized expansion coefficients
				 $\xi_{k=1,\dotsc,4}^{|i\rangle}$ (for $i=0,1$) and $\zeta_{k=1,\dotsc,60}$
				 defined by Eqs.~(\ref{eta_W}-\ref{eta_R}),
				 which correspond to the optimal readout pulses $\eta^{(W)}_{|0\rangle}(t)$, $\eta^{(W)}_{|1\rangle}(t)$ and $\eta^{(R)}(t)$
				 depicted in the right column of Fig.~2 of the main text.
				 The coefficients for the write pulses are normalized to
				 $\eta^{(W)}=\kappa$, and for the readout pulse to $\eta^{(R)} = 0.11\cdot\kappa$.
         }
\label{tbl-pulse-coefficients-holes}
\end{table*}
\end{widetext}

\end{document}